\documentclass[aip,jpm,preprint,graphicx]{revtex4-1}
\usepackage{amssymb}
\usepackage{amsmath}
\usepackage{graphicx}
\usepackage{color}

\begin{document}

\title[Models of spin-orbit coupled oligomers]{Models of spin-orbit-coupled
oligomers}
\author{G. Gligori\'c}
\author{A. Radosavljevi\'c}
\author{J. Petrovi\'c}
\author{A. Maluckov}
\author{ Lj. Had\v zievski}
\affiliation{P* group, Vin\v ca Institute of Nuclear
	Sciences, University of Belgrade, P. O. B. 522, 11001 Belgrade,
	Serbia }

\author{B. A. Malomed}
\affiliation{Department of Physical Electronics, School of Electrical Engineering,
Faculty of Engineering, and Center for Light-Matter Interaction,
Tel Aviv
University, Tel Aviv 69978, Israel\\
ITMO University, St. Petersburg 197101, Russia}

\date{\today}

\begin{abstract}
We address the stability and dynamics of eigenmodes in linearly-shaped
strings (dimers, trimers, tetramers, and pentamers) built of droplets of a
binary Bose-Einstein condensate (BEC). The binary BEC is composed of atoms
in two pseudo-spin states with attractive interactions, dressed by properly
arranged laser fields, which induce the (pseudo-) spin-orbit (SO) coupling.
We demonstrate that the SO-coupling terms help to create eigenmodes of
particular types in the strings. Dimer, trimer, and pentamer eigenmodes of
the linear system, which correspond to the zero eigenvalue (EV, alias
chemical potential) extend into the nonlinear ones, keeping an exact
analytical form, while tetramers do not admit such a continuation, because
the respective spectrum does not contain a zero EV. Stability areas of these
modes shrink with the increasing nonlinearity. Besides these modes, other
types of nonlinear states, which are produced by the continuation of their
linear counterparts corresponding to some nonzero EVs, are found in a
numerical form (including ones for the tetramer system). They are stable in
nearly entire existence regions in trimer and pentamer systems, but only in
a very small area for the tetramers. Similar results are also obtained, but
not displayed in detail, for hexa- and septamers.
\end{abstract}

\pacs{42.82.Et; 03.65.Ge; 03.75.Mn}
\maketitle

\textbf{Discrete (lattice) dynamical systems, built with intersite linear
coupling and onsite nonlinearity, are a subject of theoretical and
experimental studies in a vast research area. While the dynamics of
large-size dynamical lattices may be quite complex, in many cases insight
into their fundamental dynamical properties is provided by truncation to
small networks. In the simplest case, these are one-dimensional strings
including several lattice sites. Recently, a new paradigm of combined linear
and nonlinear dynamics was introduced in two-component Bose-Einstein
condensates (BECs), with the components which are linearly mixed, through
the first-order spatial derivatives, by the (pseudo-) spin-orbit (SO)
coupling. It has been found that the interplay of linear and nonlinear
interactions, specific to the SO-coupled BEC, gives rise to many unusual
modes. In particular, two-dimensional solitons in the free space, which are
always unstable in conventional systems, may be stabilized by the SO
coupling. This analysis produces novel results in continuous and discrete
versions of the nonlinear SO-coupled system alike. The objective of this
work is to identify stable nonlinear modes in finite strings, composed of
two, three, four, or five lattice sites (oligomers), under the combined
action of the onsite cubic self- and cross-attraction and linear SO
coupling, in the two-component BEC. In the dynamical strings considered
here, new modes are found. These include, on the one hand, linear eigenmodes
corresponding to zero eigenvalues (EVs), which can be extended without any
change in shape in the presence of nonlinearity, and, on the other hand,
nonlinear states which, in the trimer and pentamer systems, originate from
eigenmodes associated with some nonzero eigenvalues in the limit of
vanishing nonlinearity, but have their shapes different from those of the
linear counterparts. In the tetramer system, a family of nonlinear modes is
found too, but, on the contrary to those in the trimer and pentamer cases,
its stability range is very narrow. Lastly, the generality of the findings
is confirmed by the fact that similar results are obtained (but not reported
here in detail) for the oligomers built of six and seven sites.}

\section{Introduction}

The behavior of complex networks with interactions of various types between
constituents is a vast research area with diverse applications to physics
and other fields of science \cite{Sakaguchi,networks}. To address a more
particular and practically tractable class of such systems, we here study
small sets (\textit{strings}) formed of $N$ coupled pseudospinor
(two-component) droplets of atomic Bose-Einstein condensates (BECs), which
is an important physical setting \cite{droplets}, that can be realized in
the experiment by means of deep optical-lattice potentials \cite%
{deep-OL,expp}. The corresponding eigenmodes were found, and effects of the
linear spin-orbit (SO) coupling \cite{9,10,11,12} and nonlinear intra- and
inter- component interactions on their shape and stability were studied in
detail.

The background of our analysis refers to light transmission in networks
consisting of a small number ($N$) of linearly coupled fibers, or waveguides
mounted on a chip. These systems were analytically studied in special cases,
such as dimers, various forms of trimer configurations \cite{trimer,
tsiron,Buryak,tsiron1,Angelis,Aceves,Kip,Chiang} some other very specific
settings \cite{hening}. In addition, the self-trapping transition in dimers
was studied analytically and numerically, with respect to the occurrence of
a bifurcation in the stationary states of the system \cite{Eilbeck} and its
temporal dynamics \cite{tsiron}. The actual systems are formulated below in
terms of BEC, rather than arrayed optical waveguides, as the former
interpretation is more straightforward and realistic \cite{New,kett}.

The 1D SO-coupled \textit{N-mer} BEC systems with $2N$ degrees of freedom,
built as strings of $N=2,3,4,5$ adjacent BEC droplets, can be defined as a
straightforward truncation of the discrete lattice for fields $\phi
_{n}^{+,-}$, introduced in Ref. \cite{HS}, with the discrete coordinate
taking values $n=0,1~(N=2)$ , $n=0,\pm 1~(N=3)$, $n=0,\pm 1,2~(N=4)$, and $%
n=0,\pm 1,\pm 2~(N=5)$, respectively. Accordingly, the evolution of the wave
function is governed by the system of $2N$ coupled discrete Gross-Pitaevskii
equations (GPEs). The peculiarity of these systems is the double number of
degrees of freedom due to the presence of two coupled components of the
underlying spinor wave function. Moreover, the coupling constants
(parameters) may take complex values, which contributes to diversity of the
eigenmodes in these small systems, although the structure of some modes is
simplified by specific symmetries.

In contrast to convoluted situations in solid-state physics, the "synthetic"
SO coupling, induced by the appropriate laser illumination of atomic gases
in the combination with a magnetic field, can be precisely controlled in the
experiment \cite{9,10,11,12}. The link between the solid-state and
atomic-gas settings is established by mapping the electrons' spinor wave
functions onto pseudo-spinor mean-field wave functions of the binary BEC,
which is composed of atoms in two different states, ``dressed" by the laser
fields. In particular, making use of the electronic ground-state hyperfine
manifolds $5S_{1/2}$, $F=1$ in $^{87}$Rb, one can start with the set of two
hyperfine ground states, \textit{viz}., $\left\vert
F=1,m_{F}=-1\right\rangle $ and $\left\vert F=1,m_{F}=0\right\rangle $. This
pair of atomic states emulate the set of spin-up ($+$) and spin-down ($-$)
components of the electrons' wave functions. The 1D discrete model with the
intrinsic SO coupling of the Rashba type and attractive nonlinearity,
introduced in Refs. \cite{HS,New1,epjou,New3}, is adopted in the present
work to predict various steady states and their stability in strings of
tunnel-coupled BEC droplets. As mentioned above, the possibility to use deep
optically induced potentials makes such strings objects of direct relevance
to the experiment \cite{droplets,deep-OL}.

The rest of the paper is structured as follows. The model equations, their
basic properties, and a brief overview of numerical procedures employed in
the present work are introduced in Sec. II. In Sections III - V, nonlinear
modes in dimer, trimer, tetramer and pentamer systems of coupled droplets
are addressed in necessary details. The presentation in each section starts
with the analysis of the eigenvalue problem of the corresponding linear
system, without and with the SO coupling. The next step is searching for
nonlinear eigenmodes which are a direct continuation of linear ones, and the
analysis of their stability. Such nonlinear solutions are found, in an \emph{%
exact analytical form}, in the dimer, trimer, and pentamer systems, for the
modes which correspond to zero chemical potential (eigenfrequency), $\mu $,
in the linear limit, but they do not exists in the case of tetramers. Some
linear eigenmodes corresponding to $\mu \neq 0$ can be also be extended to
the nonlinear systems, but in a numerical form. Nonlinear modes of the
latter type in trimers, tetramers, and pentamers are reported in Section VI.
The paper is concluded by Section VII. In particular, in that section we
mention that results similar to those for tetramers have been obtained in
the hexamer system (another one with an even number of sites), and results
similar to what has been found for trimers and pentamers were also produced
for septamers (the next system with an odd number), although the latter
cases are not reported in the paper, to keep its size in reasonable limits.

\section{The model equations}

The fragmentation of the BEC in the deep optical lattice (OL) potential into
droplets, coupled by tunnelling across barriers separating local potential
wells, leads to the replacement of the continuous Gross-Pitaevskii equations
(GPEs), which provide for the mean-field approximation, by their discrete
counterparts. This can be performed using the tight-binding approximation as
it was elaborated in Refs. \cite{HS,New1,nashsoc} for the SO-coupled
pseudo-spinor BEC wave function, $\Phi _{n}=(\phi _{n}^{+},\phi
_{n}^{-})^{T} $, where $n$ is the discrete coordinate:
\begin{eqnarray}
i\frac{d\phi _{n}^{+}}{dt} &=&-\frac{1}{2}\left( \phi _{n-1}^{+}+\phi
_{n+1}^{+}\right) -\left( \beta |\phi _{n}^{+}|^{2}+\kappa |\phi
_{n}^{-}|^{2}\right) \phi _{n}^{+}+\lambda \left( -\phi _{n-1}^{-}+\phi
_{n+1}^{-}\right) ,  \notag \\
i\frac{d\phi _{n}^{-}}{dt} &=&-\frac{1}{2}\left( \phi _{n-1}^{-}+\phi
_{n+1}^{-}\right) -\left( \beta |\phi _{n}^{-}|^{2}+\kappa |\phi
_{n}^{+}|^{2}\right) \phi _{n}^{-}+\lambda \left( \phi _{n-1}^{+}-\phi
_{n+1}^{+}\right) .  \label{nsys}
\end{eqnarray}%
Here, components $\phi _{n}^{\pm }$ represent two different hyperfine atomic
states, $\beta >0$ and $\kappa >0$ are strengths of the self- and
cross-attraction of the two components, respectively, and $\lambda $ is the
SO-coupling strength. The case of the Manakov's nonlinearity, with $\beta
=\kappa $, will play a special role in the analysis presented below, as it
is close to the real situation in binary BECs, where the intra- and
inter-component scattering lengths are almost exactly equal \cite{Ho}. By
means of rescaling, we fix $\beta =1$.

Coupling constant $\lambda $ is taken real and positive in Eq. (\ref{nsys})
because, starting from a complex or negative one, $\lambda =|\lambda |\exp
\left( i\chi \right) $ (then, it appears as $\lambda ^{\ast }$ in the
equation for $\phi _{n}^{-}$, with the asterisk standing for the complex
conjugate), its phase can be eliminated by means of a simple phase shift of
the discrete fields, $\phi _{n}^{\pm }\rightarrow \phi _{n}^{\pm }\exp
\left( \pm i\chi /2\right) $. While Eq. (\ref{nsys}) assumes the attractive
on-site interaction, the originally repulsive sign can be transformed into
the attractive one by means of the well-known staggering transformation \cite%
{sttrf}, $\Phi _{n}=(-1)^{n}\Phi _{n}^{\ast }$.

We here consider the one-dimensional SO-coupled \textit{N-mer} BEC systems
with $2N$ degrees of freedom, built as strings of $N=2,3,4,5$ adjacent
droplets. These systems are introduced as a straightforward truncation of
the full discrete model given by Eq.(\ref{nsys}) for fields $\phi _{n}^{\pm
} $, with the discrete coordinate taking values $n=0,1(N=2)$ , $n=0,\pm
1(N=3)$, $n=0,\pm 1,2(N=4)$ and $n=0,\pm 1,\pm 2(N=5)$, respectively.
Accordingly, at $n$ which does not belong to these sets, we set $\phi
_{n}^{\pm }\equiv 0$ (for example, $\phi _{-1}^{\pm }=\phi _{2}^{\pm }\equiv
0$ in the case of $N=2$).

We have also looked for a special type of solutions which obey constraint
\begin{equation}
\phi _{n}^{+}=\pm i\phi _{n}^{-}.  \label{single}
\end{equation}%
In that case, system Eq.(\ref{nsys}) can be reduced to the following \emph{%
single} discrete nonlinear Schr\"{o}dinger equation (DNLSE):
\begin{equation}
i\frac{d\phi _{n}}{dt}=-\left( \frac{1}{2}+i\lambda \right) \phi
_{n-1}-\left( \frac{1}{2}-i\lambda \right) \phi _{n+1}-\left( 1+\kappa
\right) |\phi _{n}|^{2}\phi _{n}.  \label{singcomp}
\end{equation}%
Further, substitution $\phi _{n}\equiv \left( \frac{1+2i\lambda }{\sqrt{%
1+4\lambda ^{2}}}\right) ^{n}\varphi _{n}$ and $\tau \equiv \sqrt{1+4\lambda
^{2}}t$ transforms Eq. (\ref{singcomp}) into the standard DNLSE:
\begin{equation}
i\frac{d\varphi _{n}}{d\tau }=-\frac{1}{2}\left( \varphi _{n-1}+\varphi
_{n+1}\right) -\frac{1+\kappa }{\sqrt{1+4\lambda ^{2}}}|\varphi
_{n}|^{2}\varphi _{n}.  \label{1DNLS}
\end{equation}%
Solutions of such systems of different sizes $N$ have been studied in detail
in previous works \cite{tsiron}.

The stability of various stationary modes found below is checked by means
of the linear stability analysis (LSA) and by direct numerical simulations.
Briefly speaking, the LSA is performed for linearized (Bogoliubov - de
Gennes) equations for small perturbations added to the eigenmodes \cite%
{nashsoc,nashnum}: $\mathbf{\tilde{\phi _{n}^{\pm }}}=\mathbf{\phi _{n}^{\pm
}}+\delta \mathbf{\phi _{n}^{\pm }}$, $|\delta \mathbf{\phi _{n}^{\pm }}|\ll
|\mathbf{\phi _{n}^{\pm }}|$, where $\mathbf{\tilde{\phi _{n}^{\pm }}}$ and $%
\mathbf{\phi _{n}^{\pm }}$ denote, respectively, the perturbed discrete
solution, and their unperturbed counterparts. Perturbation eigenmodes are
looked for as $\delta \phi _{n}^{\pm }(t)=\delta \phi _{n}(t=0)^{\pm }\exp
(\rho t)$, where $\rho $ is a complex eigenvalue (EV)\ whose positive real
part, if any, implies the instability of the solution. The instability is
categorized as exponential, if the corresponding imaginary part of $\rho $
is zero, or as an oscillatory instability otherwise. The perturbed solution
is substituted into the underlying equations, which are linearized with
respect to the small perturbations. By means of a straightforward algebraic
procedure, the resulting system may be reduced to the EV problem with the
corresponding matrix \cite{nashnum}, which is then solved numerically .

To confirm the LSA predictions and fully explore the (in)stability of the
nonlinear modes, the evolution of the modes, with small random perturbations
of relative amplitude $0.01$ added to them, was numerically simulated by
means of the six-order Runge-Kutta algorithm. The numerical convergence was
checked at each step -- in particular, by monitoring the conservation of the
norm and Hamiltonian (energy).

\section{The spin--orbit-coupled dimer}

The SO-coupled dimer system with four degrees of freedom, built of two
adjacent sites, can be introduced as a straightforward truncation of the
discrete model for fields $\phi _{n}^{\pm }$, given by Eq. (\ref{nsys}),
keeping solely discrete coordinates $n=0$ and $+1$:
\begin{eqnarray}
i\frac{d\phi _{0}^{+}}{dt} &=&-\frac{1}{2}\phi _{+1}^{+}-\left( |\phi
_{0}^{+}|^{2}+\kappa |\phi _{0}^{-}|^{2}\right) \phi _{0}^{+}+\lambda \phi
_{+1}^{-},  \notag \\
i\frac{d\phi _{0}^{-}}{dt} &=&-\frac{1}{2}\phi _{+1}^{-}-\left( |\phi
_{0}^{-}|^{2}+\kappa |\phi _{0}^{+}|^{2}\right) \phi _{0}^{-}-\lambda \phi
_{+1}^{+},  \label{0di}
\end{eqnarray}%
\begin{eqnarray}
i\frac{d\phi _{+1}^{+}}{dt} &=&-\frac{1}{2}\phi _{0}^{+}-\left( |\phi
_{+1}^{+}|^{2}+\kappa |\phi _{+1}^{-}|^{2}\right) \phi _{+1}^{+}-\lambda
\phi _{0}^{-},  \notag \\
i\frac{d\phi _{+1}^{-}}{dt} &=&-\frac{1}{2}\phi _{0}^{-}-\left( |\phi
_{+1}^{-}|^{2}+\kappa |\phi _{+1}^{+}|^{2}\right) \phi _{+1}^{-}+\lambda
\phi _{0}^{+}.  \label{+1di}
\end{eqnarray}%
The conserved energy of the dimer can be written as
\begin{gather}
E_{0.+1}=-\mathrm{Re}\left[ \left( \phi _{0}^{+}\right) ^{\ast }\phi
_{+1}^{+}+\left( \phi _{0}^{-}\right) ^{\ast }\phi _{+1}^{-}\right]  \notag
\\
-\frac{1}{2}\left( |\phi _{0}^{+}|^{4}+|\phi _{+1}^{+}|^{4}+|\phi
_{0}^{-}|^{4}+|\phi _{+1}^{-}|^{4}\right) -\kappa \left( \left\vert \phi
_{0}^{+}\right\vert ^{2}\left\vert \phi _{0}^{-}\right\vert ^{2}+\left\vert
\phi _{+1}^{+}\right\vert ^{2}\left\vert \phi _{+1}^{-}\right\vert
^{2}\right)  \notag \\
+\lambda \left[ \left( \phi _{0}^{+}\right) ^{\ast }\phi _{+1}^{-}-\left(
\phi _{+1}^{+}\right) ^{\ast }\phi _{0}^{-}-\left( \phi _{0}^{-}\right)
^{\ast }\phi _{+1}^{+}+\left( \phi _{+1}^{-}\right) ^{\ast }\phi _{0}^{+}%
\right] ,  \label{E0+1}
\end{gather}%
and its norm is%
\begin{equation}
P=\sum_{+,-}\left( \left\vert \phi _{0}^{\pm }\right\vert ^{2}+\left\vert
\phi _{+1}^{\pm }\right\vert ^{2}\right) .  \label{N-dimer}
\end{equation}

Stationary solutions of system (\ref{0di})-(\ref{+1di}) with real chemical
potential $\mu $ are looked for as
\begin{equation}
\phi _{0,+1}^{\pm }=\exp \left( -i\mu t\right) u_{0,+1}^{\pm }.
\label{plane}
\end{equation}%
After the substitution of expression (\ref{plane}) in the linear version of
Eqs. (\ref{0di})-(\ref{+1di}), the following two double-degenerate EVs are
obtained:
\begin{equation}
\mu _{1,2}^{(0)}=\frac{1}{2}\sqrt{1+4\lambda ^{2}},~\mu _{3,4}^{(0)}=-\frac{1%
}{2}\sqrt{1+4\lambda ^{2}},  \label{evs}
\end{equation}%
where the upper index $(0)$ denotes that the linear system is considered.

\begin{figure}[tbp]
\centering
\includegraphics[width=14cm]{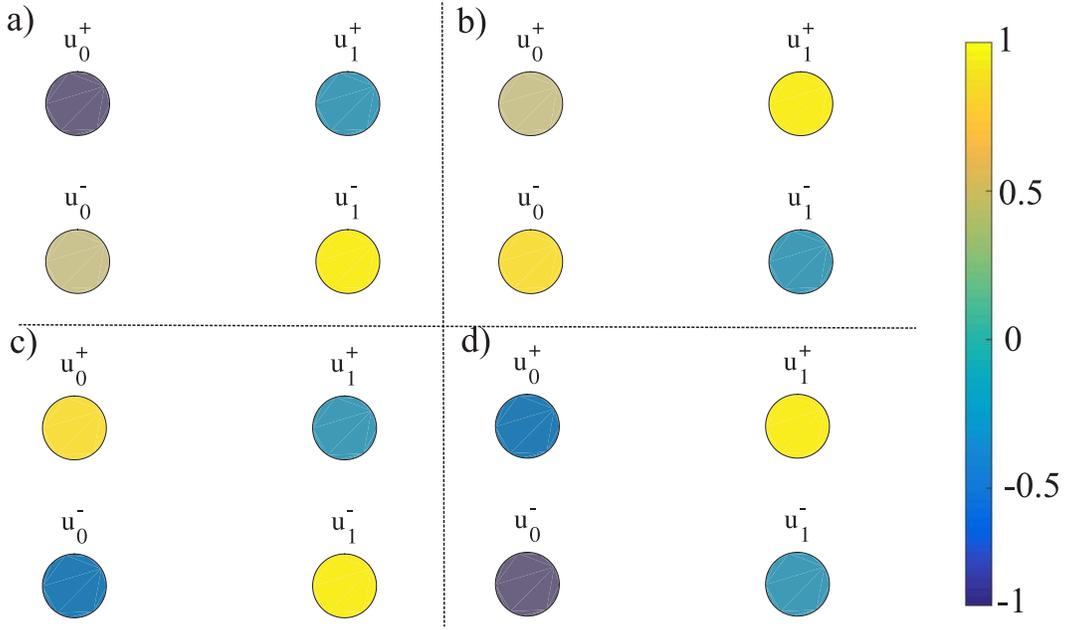}
\caption{Four eigenvectors of the linear dimer system that correspond to $%
\protect\mu^{(0)} _{1,2}$ (a,b) and $\protect\mu^{(0)}_{3,4}$ (c,d), given
by Eq. (\protect\ref{evs}), at $\protect\lambda =1$.}
\label{figevec}
\end{figure}

The corresponding eigenvectors, displayed in Fig. \ref{figevec}, are
\begin{eqnarray}
\mathbf{u_{1}} &=&\left\{ u_{0}^{+},u_{1}^{+},u_{0}^{-},u_{1}^{-}\right\}
=C\left\{ \frac{-2\lambda }{\sqrt{1+4\lambda ^{2}}},0,\frac{1}{\sqrt{%
1+4\lambda ^{2}}},1\right\} ,  \notag \\
\mathbf{u_{2}} &=&\left\{ u_{0}^{+},u_{1}^{+},u_{0}^{-},u_{1}^{-}\right\}
=C\left\{ \frac{1}{\sqrt{1+4\lambda ^{2}}},1,\frac{2\lambda }{\sqrt{%
1+4\lambda ^{2}}},0\right\} ,  \notag \\
\mathbf{u_{3}} &=&\left\{ u_{0}^{+},u_{1}^{+},u_{0}^{-},u_{1}^{-}\right\}
=C\left\{ \frac{2\lambda }{\sqrt{1+4\lambda ^{2}}},0,\frac{-1}{\sqrt{%
1+4\lambda ^{2}}},1\right\} .  \notag \\
\mathbf{u_{4}} &=&\left\{ u_{0}^{+},u_{1}^{+},u_{0}^{-},u_{1}^{-}\right\}
=C\left\{ \frac{-1}{\sqrt{1+4\lambda ^{2}}},1,\frac{-2\lambda }{\sqrt{%
1+4\lambda ^{2}}},0\right\} ,  \label{1234}
\end{eqnarray}%
where $C$ is an arbitrary constant. In the absence of the SO coupling, $%
\lambda =0$, the plus and minus components are decoupled, and the system can
be reduced to the single-component dimer.

\begin{figure}[tbp]
\centering
\includegraphics[width=14cm]{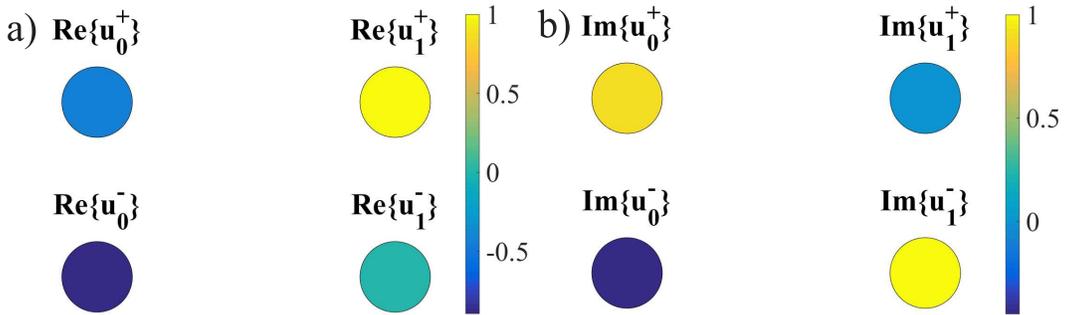}
\caption{The real and imaginary parts of the combinations of eigenvector $%
\mathbf{u_{4}}+i\mathbf{u_{3}}$ [see Eq. (\protect\ref{dimer-v})], which
provide solutions for the nonlinear dimer system at $\protect\kappa =1.5$
and $\protect\lambda =1$.}
\label{lincom}
\end{figure}

Linear eigenvectors ($\mathbf{u_{1}},\mathbf{u_{2}},\mathbf{u_{3}},\mathbf{%
u_{4}}$), given by Eq. (\ref{1234}), keep their form (shown in Fig. \ref%
{figevec}) in the presence of the nonlinearity, but solely in the
above-mentioned Manakov's case, when the intra- and inter-component
nonlinearity coefficients are equal, $\kappa =1$. This fact may be
considered as a manifestation of the principle that extension of linear
modes into nonlinear ones is facilitated if nonlinear terms in the system
obey the same symmetry as linear ones, see, e.g. Ref. \cite{New3}. The
corresponding EVs of the chemical potential, including the nonlinear shift,
are
\begin{equation}
\mu =-|C|^{2}\pm \frac{1}{2}\sqrt{1+4\lambda ^{2}},  \label{nmu}
\end{equation}%
cf. Eqs. (\ref{evs}). Varying $|C|^{2}$, one can tune the eigenfrequencies
in the nonlinear case as per Eq. (\ref{nmu}), cf. Eq. (\ref{evs}). The norm
of the nonlinear modes depends on $\lambda $, and can be found as
\begin{equation}
P=-2\mu \pm \sqrt{1+4\lambda ^{2}}.  \label{npo}
\end{equation}

To analyze the stability of the stationary nonlinear solutions, we apply the
LSA, adding small arbitrary complex-valued perturbation $\delta \phi
_{0,1}^{\pm }$ to the stationary modes. The perturbed solution is
substituted into Eqs. (\ref{0di})-(\ref{+1di}), which are then linearized
with respect to the small perturbations. The resulting linear system is then
reduced to the EV problem with matrix

\begin{equation}
\left(
\begin{array}{cccccccc}
\rho & 0 & 0 & 0 & -\mu -|C|^{2} & -1/2 & 0 & \lambda \\
0 & \rho & 0 & 0 & -\frac{1}{2} & -\mu -|C|^{2} & -\lambda & 0 \\
0 & 0 & \rho & 0 & 0 & -\lambda & -\mu -|C|^{2} & -\frac{1}{2} \\
0 & 0 & 0 & \rho & \lambda & 0 & -\frac{1}{2} & -\mu -|C|^{2} \\
\mu +\frac{12\lambda ^{2}+1}{1+4\lambda ^{2}}|C|^{2} & \frac{1}{2} & \frac{%
-4\lambda }{1+4\lambda ^{2}}|C|^{2} & -\lambda & \rho & 0 & 0 & 0 \\
\frac{1}{2} & \mu +|C|^{2} & \lambda & 0 & 0 & \rho & 0 & 0 \\
\frac{-4\lambda }{1+4\lambda ^{2}}|C|^{2} & \lambda & \mu +\frac{4\lambda
^{2}+3}{1+4\lambda ^{2}}|C|^{2} & \frac{1}{2} & 0 & 0 & \rho & 0 \\
-\lambda & 0 & \frac{1}{2} & \mu +3|C|^{2} & 0 & 0 & 0 & \rho%
\end{array}%
\right) .
\end{equation}%
Eigenvalues of this matrix can be found in an exact form:
\begin{gather}
\rho _{1,2,3,4}=0,  \notag \\
\rho _{5,6}=\pm i\sqrt{1+4\lambda ^{2}},  \label{eve} \\
\rho _{7,8}=\pm \left( 1+4\lambda ^{2}\right) ^{1/4}\sqrt{2\left( \mu -\sqrt{%
1+4\lambda ^{2}}\right) },  \notag
\end{gather}%
where $|C|^{2}$ is substituted by $-\mu \pm \frac{1}{2}\sqrt{1+4\lambda ^{2}}
$, as per Eq. (\ref{nmu}).

As it follows from Eq. (\ref{nmu}), the present stationary dimer solutions
exist in the region of $-\infty <\mu \leq \frac{1}{2}\sqrt{1+4\lambda ^{2}}$%
, hence all EVs (\ref{eve}) have zero real parts, i.e., \emph{all} the
stationary solutions are \emph{stable}. Direct simulations corroborate this
conclusion, by displaying stable evolution of perturbed stationary solutions
(not shown here in detail).

While each eigenvector from set (\ref{1234}) does not feature any intrinsic
symmetry, linear combinations of the eigenvectors, taken as
\begin{equation}
\left( \mathbf{v}_{12}\right) _{\pm }=\mathbf{u_{2}}\pm i\mathbf{u_{1}}%
,~\left( \mathbf{v}_{34}\right) _{\pm }=\mathbf{u_{4}}\pm i\mathbf{u_{3}},
\label{dimer-v}
\end{equation}%
produce solutions which obey the symmetry restriction defined by Eq. (\ref%
{single}). As said above, it corresponds to the special case when the system
is reduced to the single-component DNLSE (\ref{singcomp}), with two sites.
Unlike the eigenvectors presented by the individual eigenmodes, these
combinations, shown in Fig. \ref{lincom}, extend as nonlinear states even in
the case of the non-Manakov's nonlinearity, $\kappa \neq 1$, with the
nonlinearly shifted eigenfrequencies $\mu =-(1+\kappa )|C|^{2}\pm \frac{1}{2}%
\sqrt{1+4\lambda ^{2}}$[note that sign $\pm $ here has the same meaning as
in Eq. (\ref{nmu}), but not as in Eq. (\ref{dimer-v})]. The norm of the mode
can be found as $P=-2/(1+\kappa )(-2\mu \pm \sqrt{1+4\lambda ^{2}})$. The
respective LSA and direct simulations confirm the stability of combined
modes (\ref{dimer-v}) in their existence region.

It is relevant to mention that the availability of the exact solutions for
all the linear eigenmodes, as well as for some of the nonlinear ones, makes
the use of the well-known method of the continuation from the anticontinuum
limit \cite{anti} (which corresponds to uncoupled lattices) unnecessary in
the present case, both for the dimers and higher-order oligomers considered
below.

\section{The spin--orbit-coupled trimer}

The SO-coupled trimer string can be realized by adding an extra site $n=-1$
to the dimer. In this case, the dynamics is described by the following
system of equations:
\begin{eqnarray}
i\frac{d\phi _{-1}^{+}}{dt} &=&-\frac{1}{2}\phi _{0}^{+}-\left( |\phi
_{-1}^{+}|^{2}+\kappa |\phi _{-1}^{-}|^{2}\right) \phi _{-1}^{+}+\lambda
\phi _{0}^{-},  \notag \\
i\frac{d\phi _{-1}^{-}}{dt} &=&-\frac{1}{2}\phi _{0}^{-}-\left( |\phi
_{-1}^{-}|^{2}+\kappa |\phi _{-1}^{+}|^{2}\right) \phi _{-1}^{-}-\lambda
\phi _{0}^{+},  \label{-1}
\end{eqnarray}%
\begin{eqnarray}
i\frac{d\phi _{0}^{+}}{dt} &=&-\left( \frac{1}{2}\phi _{-1}^{+}+\frac{1}{2}%
\phi _{+1}^{+}\right) -\left( |\phi _{0}^{+}|^{2}+\kappa |\phi
_{0}^{-}|^{2}\right) \phi _{0}^{+}+\lambda \left( -\phi _{-1}^{-}+\phi
_{+1}^{-}\right) ,  \notag \\
i\frac{d\phi _{0}^{-}}{dt} &=&-\left( \frac{1}{2}\phi _{-1}^{-}+\frac{1}{2}%
\phi _{+1}^{-}\right) -\left( |\phi _{0}^{-}|^{2}+\kappa |\phi
_{0}^{+}|^{2}\right) \phi _{0}^{-}+\lambda \left( \phi _{-1}^{+}-\phi
_{+1}^{+}\right) ,  \label{0}
\end{eqnarray}%
\begin{eqnarray}
i\frac{d\phi _{+1}^{+}}{dt} &=&-\frac{1}{2}\phi _{0}^{+}-\left( |\phi
_{+1}^{+}|^{2}+\kappa |\phi _{+1}^{-}|^{2}\right) \phi _{+1}^{+}-\lambda
\phi _{0}^{-},  \notag \\
i\frac{d\phi _{+1}^{-}}{dt} &=&-\frac{1}{2}\phi _{0}^{-}-\left( |\phi
_{+1}^{-}|^{2}+\kappa |\phi _{+1}^{+}|^{2}\right) \phi _{+1}^{-}+\lambda
\phi _{0}^{+}.  \label{+1}
\end{eqnarray}

Equations (\ref{-1})-(\ref{+1}) conserve the corresponding energy,
\begin{gather}
E_{-1,0.+1}=-\mathrm{Re}\left[ \left( \phi _{-1}^{+}\right) ^{\ast }\phi
_{0}^{+}+\left( \phi _{0}^{+}\right) ^{\ast }\phi _{+1}^{+}+\left( \phi
_{-1}^{-}\right) ^{\ast }\phi _{0}^{-}+\left( \phi _{0}^{-}\right) ^{\ast
}\phi _{+1}^{-}\right]  \notag \\
-\frac{1}{2}\left( |\phi _{-1}^{+}|^{4}+|\phi _{0}^{+}|^{4}+|\phi
_{+1}^{+}|^{4}+|\phi _{-1}^{-}|^{4}+|\phi _{0}^{-}|^{4}+|\phi
_{+1}^{-}|^{4}\right) -\kappa \left( \left\vert \phi _{-1}^{+}\right\vert
^{2}\left\vert \phi _{-1}^{-}\right\vert ^{2}+\left\vert \phi
_{0}^{+}\right\vert ^{2}\left\vert \phi _{0}^{-}\right\vert ^{2}+\left\vert
\phi _{+1}^{+}\right\vert ^{2}\left\vert \phi _{+1}^{-}\right\vert
^{2}\right)  \notag \\
+\lambda \left[ \left( \phi _{-1}^{+}\right) ^{\ast }\phi _{0}^{-}-\left(
\phi _{0}^{+}\right) ^{\ast }\left( \phi _{-1}^{-}-\phi _{+1}^{-}\right)
-\left( \phi _{+1}^{+}\right) ^{\ast }\phi _{0}^{-}-\left( \phi
_{-1}^{-}\right) ^{\ast }\phi _{0}^{+}+\left( \phi _{0}^{-}\right) ^{\ast
}\left( \phi _{-1}^{+}-\phi _{+1}^{+}\right) +\left( \phi _{+1}^{-}\right)
^{\ast }\phi _{0}^{+}\right] ,  \label{E-10+1}
\end{gather}%
and the total norm,
\begin{equation}
P=\sum_{+,-}\left( \left\vert \phi _{-1}^{\pm }\right\vert ^{2}+\left\vert
\phi _{0}^{\pm }\right\vert ^{2}+\left\vert \phi _{+1}^{\pm }\right\vert
^{2}\right) .  \label{N-trimer}
\end{equation}

Assuming the usual form of stationary solutions,
\begin{equation}
\phi _{-1,0,+1}^{\pm }=\exp \left( -i\mu t\right) u_{-1,0,+1}^{\pm },
\label{plane3}
\end{equation}%
and substituting this in the linearized version of Eqs. (\ref{-1})-(\ref{+1}%
), the following three double-degenerate EVs are obtained:
\begin{equation}
\mu _{1,2}^{(0)}=0,~\mu _{3,4}^{(0)}=\sqrt{\frac{1}{2}+2\lambda ^{2}},~\mu
_{5,6}^{(0)}=-\sqrt{\frac{1}{2}+2\lambda ^{2}}.  \label{evs3}
\end{equation}

\begin{figure}[tbp]
\centering
\includegraphics[width=10cm]{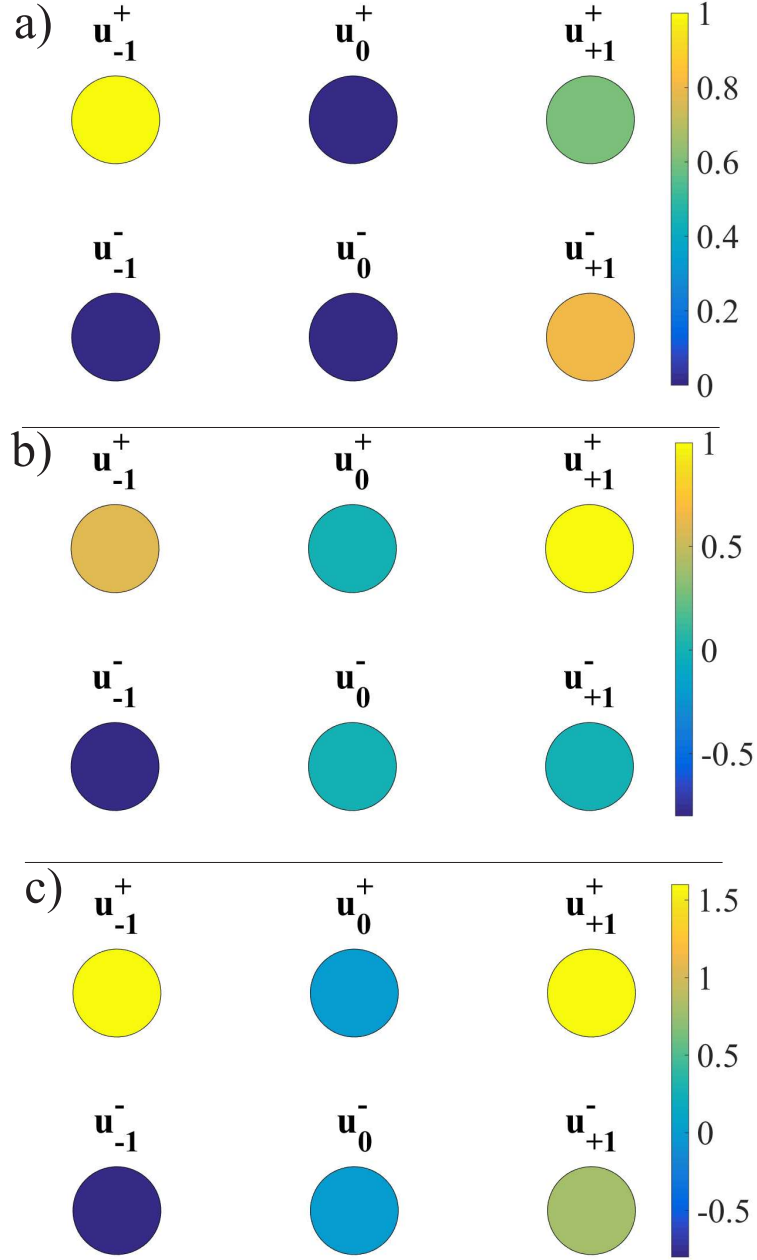}
\caption{(a,b) Eigenvectors of the linearized trimer system that correspond
to $\protect\mu _{1,2}^{(0)}=0$, and (c) their linear combination, $\mathbf{%
u_{1}}+\mathbf{u_{2}}$, at $\protect\lambda =1$.}
\label{evec3}
\end{figure}

The eigenvectors that correspond to $\mu _{1,2}^{(0)}=0$ are
\begin{eqnarray}
\mathbf{u_{1}} &=&\left\{
u_{-1}^{+},u_{0}^{+},u_{1}^{+},u_{-1}^{-},u_{0}^{-},u_{1}^{-}\right\}
=C\left\{ 1,0,\frac{4\lambda ^{2}-1}{1+4\lambda ^{2}},0,0,\frac{4\lambda }{%
1+4\lambda ^{2}}\right\} ,  \notag \\
\mathbf{u_{2}} &=&\left\{
u_{-1}^{+},u_{0}^{+},u_{1}^{+},u_{-1}^{-},u_{0}^{-},u_{1}^{-}\right\}
=C\left\{ \frac{4\lambda ^{2}-1}{1+4\lambda ^{2}},0,1,-\frac{4\lambda }{%
1+4\lambda ^{2}},0,0\right\} ,  \label{trimer_mu=0}
\end{eqnarray}%
see Fig. \ref{evec3}. Note that these eigenstates do not feature any
symmetry. On the other hand, solutions of the linearized version system (\ref%
{-1})-(\ref{+1}), obeying symmetry restrictions $u_{-1}^{+}=u_{+1}^{+}$, $%
u_{-1}^{-}=-u_{+1}^{-}$ and $u_{0}^{-}=0$, can be obtained as the linear
combination of these two eigenvectors, $\mathbf{u_{1}}+\mathbf{u_{2}}$, see
Fig. \ref{evec3}(c).

Eigenmodes (\ref{trimer_mu=0}), as well as their combination $\mathbf{u_{1}}+%
\mathbf{u_{2}}$, remain valid solutions of the nonlinear trimer system, but
only in the above-mentioned Manakov's case, $\kappa =1$, cf. the similar
result obtained above for the dimers. The respective nonlinear shift of the
chemical potential, $\mu $, and the total norm,
\begin{equation}
P=-2\mu ,  \label{Ptri}
\end{equation}%
do not depend on the SO-coupling parameter $\lambda $ (recall we consider
the solution which has $\mu =0$ in the linear limit). The (dynamically
stable) nonlinear extension of eigenvectors that correspond to $~\mu
_{5,6}^{(0)}$ , see Eq. (\ref{evs3}), is reported separately in Section VI,
as it can be found solely in a numerical form.

\begin{figure}[tbp]
\centering
\includegraphics[width=14cm]{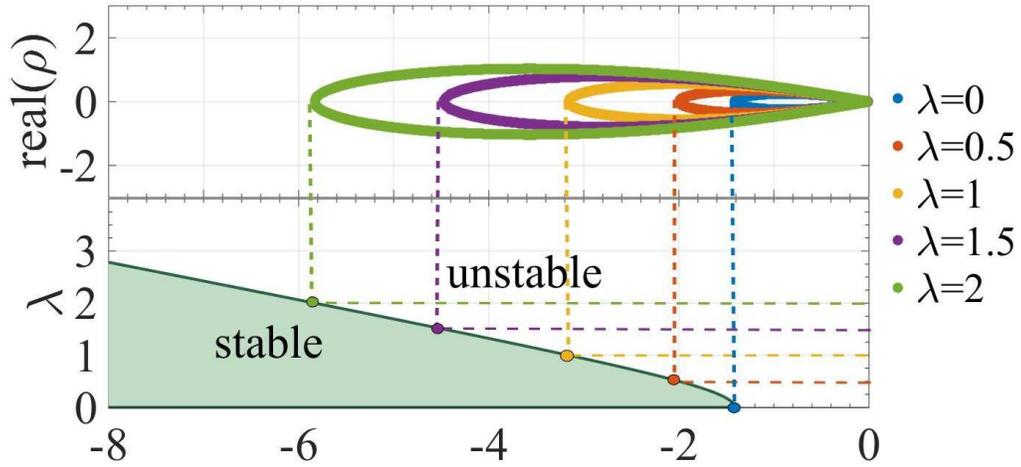}
\caption{(Top) Numerically obtained real parts of stability eigenvalues $%
\protect\rho $ for the nonlinear trimer states given by Eq. (\protect\ref%
{trimer_mu=0}), vs. $\protect\mu $ for solutions $\mathbf{u_{1}}$, $\mathbf{%
u_{2}}$ from Eq. (\protect\ref{trimer_mu=0}) at different values of $\protect%
\lambda $. (Bottom) The corresponding stability diagram, in the $\left(
\protect\mu ,\protect\lambda \right) $ plane, with the stability boundary
given by Eq. (\protect\ref{boundary}).}
\label{trimstab}
\end{figure}

\begin{figure}[tbp]
\centering
\includegraphics[width=14cm]{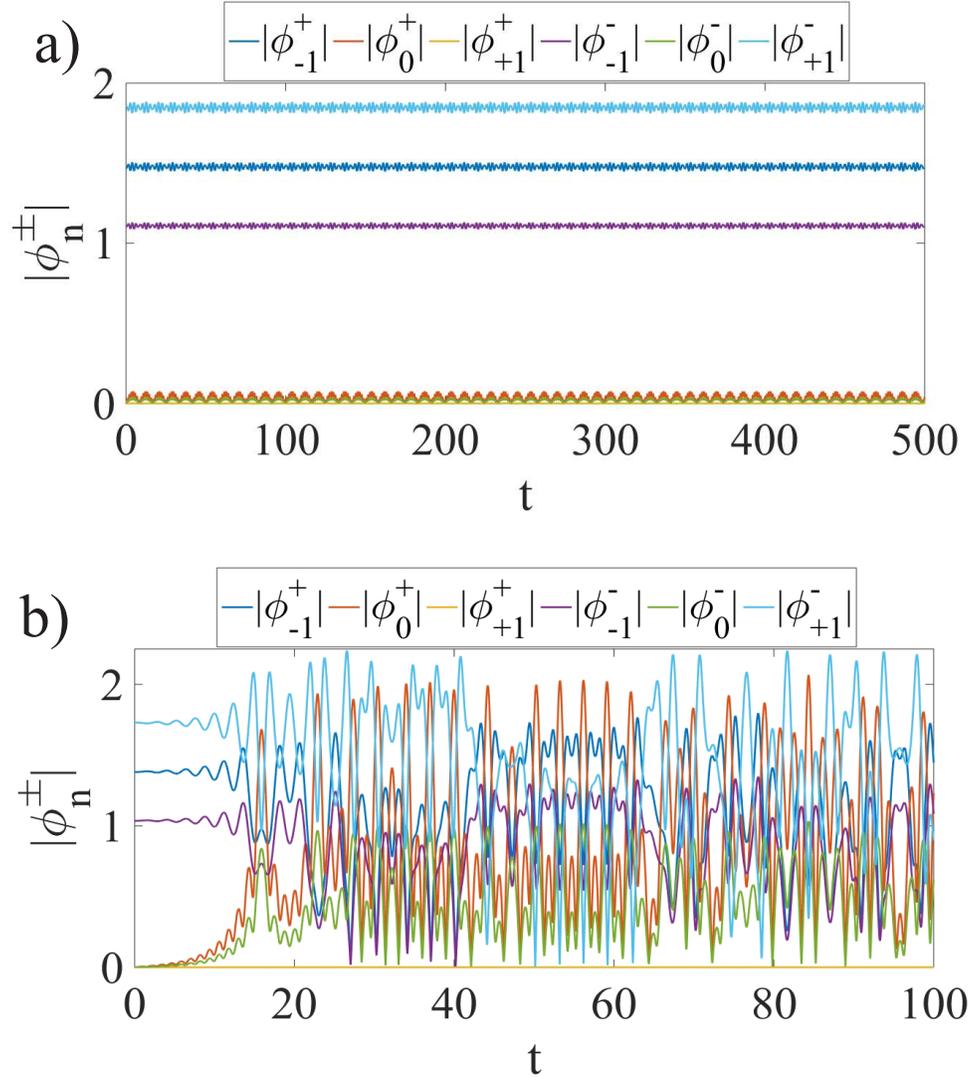}
\caption{(a) Amplitudes of a perturbed stable solution $\mathbf{u_{1}}$ [see
Eq. (\protect\ref{trimer_mu=0})], produced by simulations of the nonlinear
trimer at $\protect\mu =-3.5$ and $\protect\lambda =1$. (b) The evolution of
unstable trimer at $\protect\mu =-3$ and $\protect\lambda =1$. }
\label{trimevol}
\end{figure}

The stability of these solutions of the nonlinear system with the Manakov's
nonlinearity, $\kappa =1$, was analyzed by means of the LSA, following the
same procedure as in the case of the dimer. An instability, i.e., positive
real parts of the corresponding stability EVs, are found, in an analytical
form (confirmed by the numerical calculation), at
\begin{equation}
\mu >\mu _{\mathrm{cr}}^{\mathrm{(trimer)}}\equiv -2\sqrt{\frac{1}{2}%
+2\lambda ^{2}},  \label{boundary}
\end{equation}%
see Fig. \ref{trimstab}. Direct simulations corroborate the stability and
instability of the trimer modes, as predicted by Eq. (\ref{boundary}), see
Fig. \ref{trimevol}.

\begin{figure}[tbp]
\centering
\includegraphics[width=14cm]{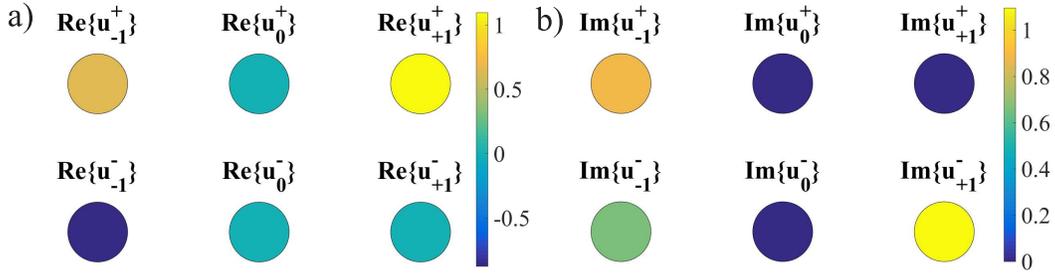}
\caption{The real and imaginary parts of the trimer eigenvector found as the
solution of the nonlinear system at $\protect\mu =-3$, $\protect\kappa =1.5$%
, and $\protect\lambda =1$.}
\label{3com}
\end{figure}

If we are looking for a solution subject to constraint (\ref{single}), the
SO-coupled trimer system can be reduced to the single-component DNLSE with
three sites:
\begin{eqnarray}
i\frac{d\phi _{-1}}{dt} &=&-\left( \frac{1}{2}-i\lambda \right) \phi
_{1}-\left( 1+\kappa \right) |\phi _{-1}|^{2}\phi _{-1},  \notag \\
i\frac{d\phi _{0}}{dt} &=&-\left( \frac{1}{2}+i\lambda \right) \phi
_{-1}-\left( \frac{1}{2}-i\lambda \right) \phi _{1}-\left( 1+\kappa \right)
|\phi _{0}|^{2}\phi _{0},  \label{3mersingcomp} \\
i\frac{d\phi _{1}}{dt} &=&-\left( \frac{1}{2}+i\lambda \right) \phi
_{0}-\left( 1+\kappa \right) |\phi _{1}|^{2}\phi _{1}.  \notag
\end{eqnarray}%
In the linear limit, Eq. (\ref{3mersingcomp}) yields the same EVs as given
by Eq. (\ref{evs3}). The eigenvector that corresponds to EV $\mu ^{(0)}=0$
is (see Fig. \ref{3com})
\begin{equation}
\left\{
u_{-1}^{+},u_{0}^{+},u_{1}^{+},u_{-1}^{-},u_{0}^{-},u_{1}^{-}\right\}
=C\left\{ \frac{\left( 4\lambda ^{2}-1\right) +4i\lambda }{1+4\lambda ^{2}}%
,0,1,\frac{-4\lambda +i\left( 4\lambda ^{2}-1\right) }{1+4\lambda ^{2}}%
,0,i\right\} .  \label{b=/=k}
\end{equation}%
It persists as an exact solution of the nonlinear system, even if $\kappa
\neq 1$, with the respective nonlinear shift of the chemical potential, $\mu
$, and the total norm $P=-4\mu /(1+\kappa )$, which do not depend on the
SO-coupling strength, $\lambda $.

The stability of the nonlinear solution (\ref{b=/=k})\ was analyzed by means
of the LSA. The following EVs have been thus obtained, in the analytical
form:
\begin{eqnarray}
\rho _{1,2} &=&0,  \notag \\
\rho _{3,4} &=&\pm i\frac{\sqrt{1+4\lambda ^{2}+\mu ^{2}-\mu \sqrt{\mu
^{2}-2(1+4\lambda ^{2})}}}{\sqrt{2}},  \label{eve3mer} \\
\rho _{5,6} &=&\pm i\frac{\sqrt{1+4\lambda ^{2}+\mu ^{2}+\mu \sqrt{\mu
^{2}-2(1+4\lambda ^{2})}}}{\sqrt{2}}.  \notag
\end{eqnarray}%
They feature an instability in the region identical to that defined by Eq. (%
\ref{boundary}), i.e., the same as displayed for $\mathbf{u_{1}}$ and $%
\mathbf{u_{2}}$ in Fig. \ref{trimstab}, with the difference that this
solution exists at arbitrary $\kappa $. Direct simulations confirm this
stability prediction.

\section{The spin--orbit-coupled tetramer and pentamer}

The tetramer string, with four sites (eight degrees of freedom), can be
built by adding a new site to the trimer. Following the same procedure as in
the cases of the dimer and trimer, we find four double-degenerate EVs from
the respective linearized system:
\begin{eqnarray}
\mu _{1,2}^{(0)} &=&-\frac{1}{4}(\sqrt{5}-1)\sqrt{1+4\lambda ^{2}},~\mu
_{3,4}^{(0)}=\frac{1}{4}(\sqrt{5}-1)\sqrt{1+4\lambda ^{2}},  \notag \\
\mu _{5,6}^{(0)} &=&-\frac{1}{4}(\sqrt{5}+1)\sqrt{1+4\lambda ^{2}},~\mu
_{7,8}^{(0)}=\frac{1}{4}(\sqrt{5}+1)\sqrt{1+4\lambda ^{2}}.  \label{evs4}
\end{eqnarray}%
The corresponding eigenvectors have been found too. Because the set of EVs (%
\ref{evs4}) does not include $\mu =0$, the tetramer system does not produce
any analytically available nonlinear mode, unlike the trimer states given by
Eqs. (\ref{trimer_mu=0}) and (\ref{b=/=k}). A numerically found (chiefly,
unstable) nonlinear mode, originating from $\mu _{5,6}^{(0)}$ in Eq. (\ref%
{evs4}), is reported in Section VI.

\begin{figure}[tbp]
\centering
\includegraphics[width=10cm]{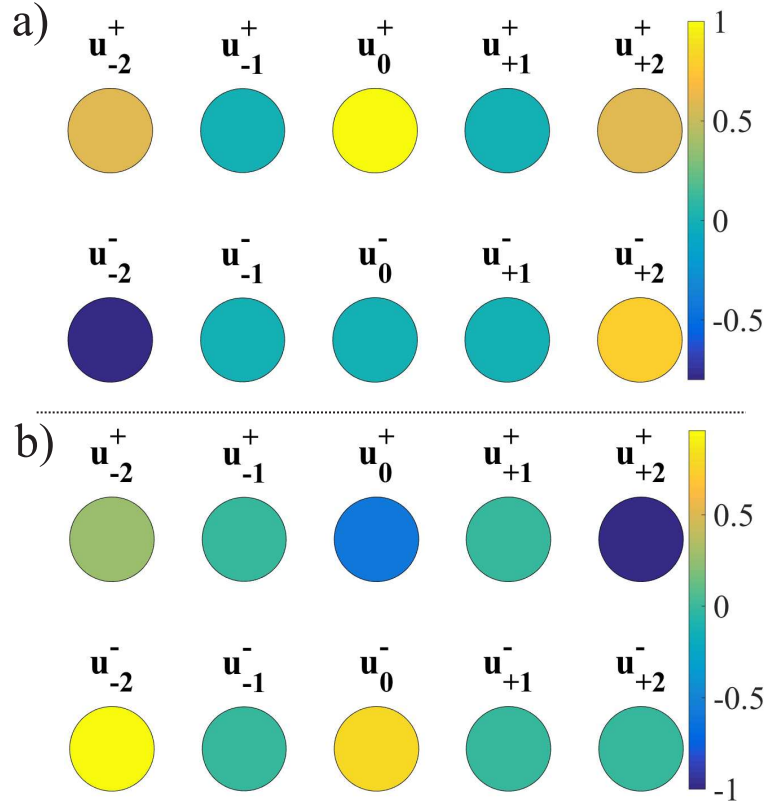}
\caption{Pentamer's eigenvectors (\protect\ref{5_mu=0}) that correspond to $%
\protect\mu _{1,2}^{(0)}=0$, at $\protect\lambda =1$.}
\label{evec5}
\end{figure}

The \textit{pentamer} string, with ten degrees of freedom, is built as a
chain of five sites, at $n={-2,-1,0,1,2}$. Following the same procedure as
above, five double-degenerate EV branches are found for the linearized
pentamer:
\begin{eqnarray}
\mu _{1,2}^{(0)} &=&0,  \notag \\
\mu _{3,4}^{(0)} &=&\frac{1}{2}\sqrt{1+4\lambda ^{2}},  \notag \\
\mu _{5,6}^{(0)} &=&-\frac{1}{2}\sqrt{1+4\lambda ^{2}},  \notag \\
\mu _{7,8}^{(0)} &=&\frac{\sqrt{3}}{2}\sqrt{1+4\lambda ^{2}},  \notag \\
\mu _{9,10}^{(0)} &=&-\frac{\sqrt{3}}{2}\sqrt{1+4\lambda ^{2}}.  \label{evs5}
\end{eqnarray}%
Eigenvectors that correspond to $\mu _{1,2}^{(0)}=0$ are
\begin{eqnarray}
\mathbf{u_{1}} &=&\left\{
u_{-2}^{+},u_{-1}^{+},u_{0}^{+},u_{1}^{+},u_{2}^{+},u_{-2}^{-},u_{-1}^{-},u_{0}^{-},u_{1}^{-},u_{2}^{-}\right\} =
\notag \\
&=&C\left\{ \frac{4\lambda ^{2}-1}{1+4\lambda ^{2}},0,1,0,\frac{4\lambda
^{2}-1}{1+4\lambda ^{2}},-\frac{4\lambda }{1+4\lambda ^{2}},0,0,0,\frac{%
4\lambda }{1+4\lambda ^{2}}\right\} ,  \notag \\
\mathbf{u_{2}} &=&\left\{
u_{-2}^{+},u_{-1}^{+},u_{0}^{+},u_{1}^{+},u_{2}^{+},u_{-2}^{-},u_{-1}^{-},u_{0}^{-},u_{1}^{-},u_{2}^{-}\right\} =
\notag \\
&=&C\left\{ -\frac{(4\lambda ^{2}-1)^{2}-16\lambda ^{2}}{(1+4\lambda
^{2})^{2}},0,-\frac{4\lambda ^{2}-1}{1+4\lambda ^{2}},0,-1,\frac{8\lambda
(4\lambda ^{2}-1)}{(1+4\lambda ^{2})^{2}},0,\frac{4\lambda }{1+4\lambda ^{2}}%
,0,0\right\} ,  \label{5_mu=0}
\end{eqnarray}%
see Fig. \ref{evec5}. Note that eigenvector $\mathbf{u_{1}}$ obeys
constraints $u_{-n}^{+}=u_{n}^{+}$, $u_{-n}^{-}=-u_{n}^{-}$, and $%
u_{0}^{-}=0 $.

\begin{figure}[tbp]
\centering
\includegraphics[width=14cm]{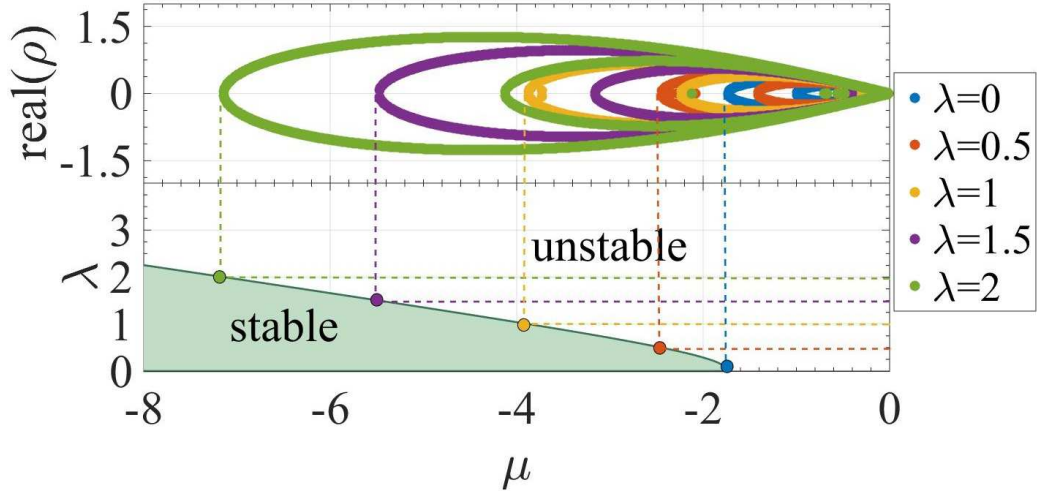}
\caption{(Top) Numerically obtained real parts of stability eigenvalues $%
\protect\rho $ vs. $\protect\mu $ for nonlinear pentamer solutions $\mathbf{%
u_{1}}$ and $\mathbf{u_{2}}$, given by Eq. (\protect\ref{5_mu=0}). (Bottom)
The corresponding stability diagram, with the stability boundary given by
Eq. (\protect\ref{penta-boundary}).}
\label{5merstab}
\end{figure}

Like in the trimer case, the eigenvectors that correspond to $\mu
_{1,2}^{(0)}=0$ extend as solutions of the full nonlinear system, but solely
under the Manakov's condition, $\kappa =1$. Once again, their nonlinear
frequency shift $\mu $ and total norm,
\begin{equation}
P=-3\mu ,  \label{Ppenta}
\end{equation}%
do not depend on $\lambda $, cf. a similar relation (\ref{Ptri}) for the
trimers. The LSA, implemented for these nonlinear states, predicts
instability at
\begin{equation}
\mu >\mu _{\mathrm{cr}}^{\mathrm{(pentamer)}}\equiv -\sqrt{3+12\lambda ^{2}},
\label{penta-boundary}
\end{equation}%
and stability at $\mu <\mu _{\mathrm{cr}}^{\mathrm{(pentamer)}}$ [cf. Eq. (%
\ref{boundary})], see Fig. \ref{5merstab}. This prediction has been
confirmed by direct simulations (not shown here in detail, as the results
are quite similar to those displayed above for stable and unstable trimers
in Fig. \ref{trimevol}).

\begin{figure}[tbp]
\centering
\includegraphics[width=10cm]{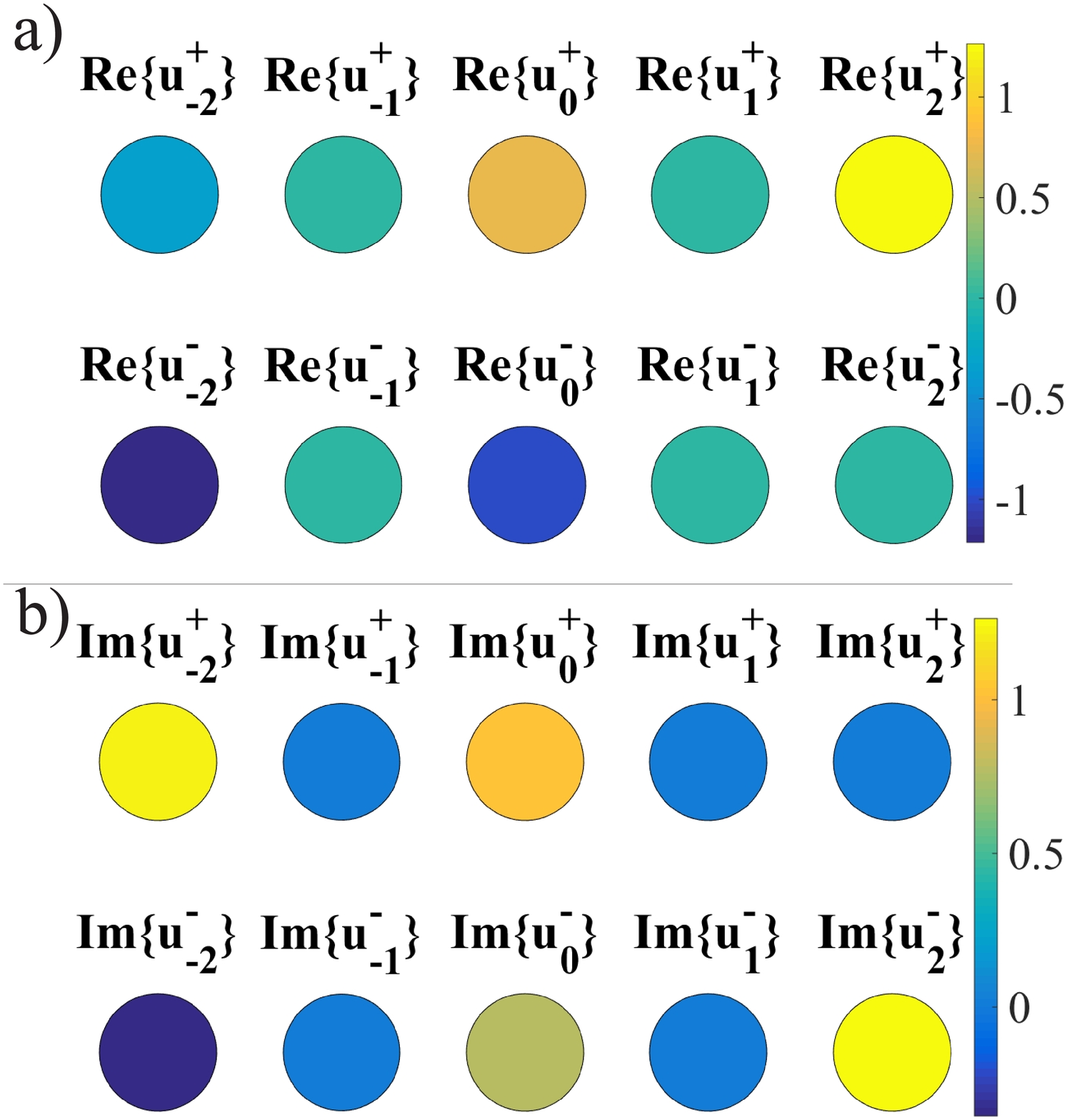}
\caption{The real and imaginary parts of eigenvector (\protect\ref{5_mu=0})
of the nonlinear pentamer system, that satisfies constraint (\protect\ref%
{single}) for $\protect\mu =-4$, $\protect\kappa =1.5$, and $\protect\lambda %
=1$.}
\label{evecsig5}
\end{figure}

Looking for solutions that satisfy constraint (\ref{single}), the SO-coupled
pentamer can be reduced to the single-component chain with five sites. The
respective linearized system yields the same eigenvalues as produced above
in Eq. (\ref{evs5}). The eigenvector that corresponds to $\mu ^{(0)}=0$ is
\begin{gather}
\left\{
u_{-2}^{+},u_{-1}^{+},u_{0}^{+},u_{1}^{+},u_{2}^{+},u_{-2}^{-},u_{-1}^{-},u_{0}^{-},u_{1}^{-},u_{2}^{-}\right\} =
\notag \\
=C\left\{ \frac{\left( 2\lambda +i\right) ^{4}}{\left( 1+4\lambda
^{2}\right) ^{2}},0,\frac{\left( 2\lambda +i\right) ^{2}}{1+4\lambda ^{2}}%
,0,1,\pm i\frac{\left( 2\lambda +i\right) ^{4}}{\left( 1+4\lambda
^{2}\right) ^{2}},0,\pm i\frac{\left( 2\lambda +i\right) ^{2}}{1+4\lambda
^{2}},0,\pm i\right\} ,  \label{5_mu=0_reduced}
\end{gather}%
see Fig. \ref{evecsig5}. This eigenvector keeps its form in the presence of
the nonlinearity, irrespective of the value of the nonlinear parameter $%
\kappa $ (i.e., in the general non-Manakov case). The corresponding
nonlinear shift of the chemical potential, $\mu $, and the total norm,
\begin{equation}
P=-6\mu /(1+\kappa ),  \label{Ppenta2}
\end{equation}%
do not depend on the SO-coupling strength, $\lambda $, similar to what was
found above for other solutions subject to constraint (\ref{single}), cf.
Eqs. (\ref{Ptri}) and (\ref{Ppenta}). The LSA demonstrates that the
nonlinear solutions, which originate from the linear eigenvector (\ref%
{5_mu=0_reduced}), precisely share the stability properties with their
counterparts, $\mathbf{u_{1}}$ and $\mathbf{u_{2}}$, see Fig. \ref{5merstab}%
. This prediction has been confirmed by direct simulations.

The linear mode corresponding to $\mu _{9,10}^{(0)}$ in Eq. (\ref{evs5}) can
numerically extended into a stable nonlinear state, as shown in the
following section.

\section{Nonlinear modes originating from nonzero eigenvalues in the linear
limit}

In the previous section we have shown that the linear eigenmodes associated
with eigenvalue $\mu ^{(0)}=0$ persist in nonlinear trimer and pentamer
systems without any change of the shape. The increase of the nonlinearity
strength only affects their stability, in the sense that, above a certain
value of the strength, these modes become unstable. The other property of
these modes is the independence of the norm on the SO-coupling parameter.

\begin{figure}[tbp]
\centering\includegraphics[width=14cm]{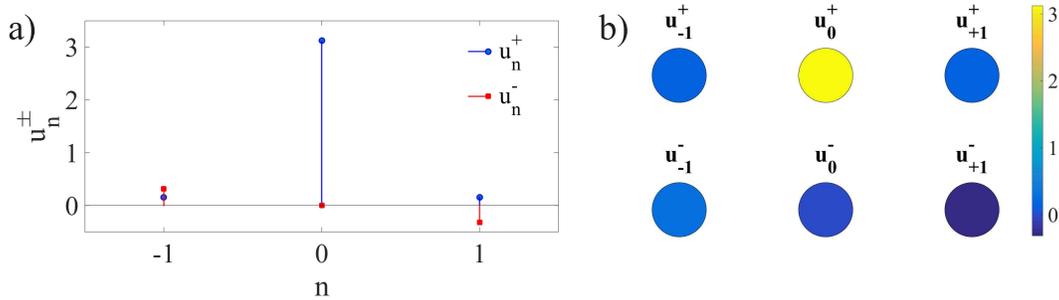}
\caption{The numerically found stable solution for the nonlinear trimer
mode, which, in the linear limit, originates from the linear eigenmode
corresponding to $\protect\mu _{5,6}^{(0)}$ in Eq. (\protect\ref{evs3}). It
is obtained for $\protect\mu =-10$, $\protect\kappa =0.5$, and $\protect%
\lambda =1$. (a) The amplitude profile of the solution. (b) Its schematic
structure.}
\label{nl3mer}
\end{figure}

\begin{figure}[tbp]
\centering
\includegraphics[width=14cm]{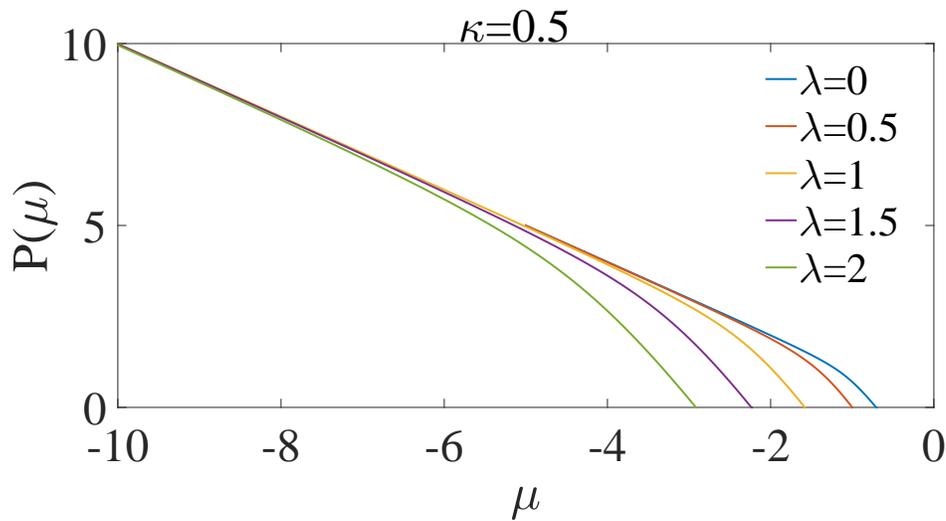}
\caption{$P(\protect\mu )$ curves for the families of nonlinear trimer
modes, originating from $\protect\mu _{5,6}^{(0)}$ in Eq. (\protect\ref{evs3}%
), at different values of $\protect\lambda $, and $\protect\kappa =0.5$.}
\label{Pm3mer}
\end{figure}

\begin{figure}[tbp]
\centering
\includegraphics[width=14cm]{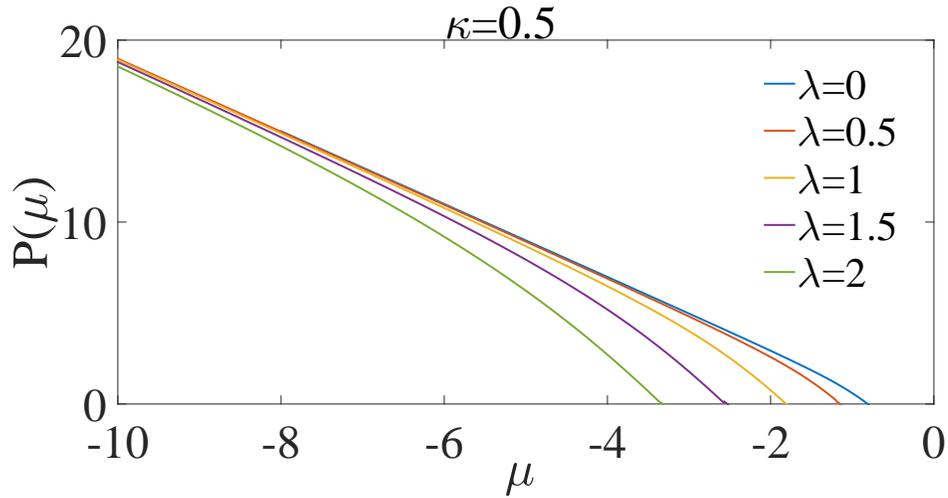}
\caption{$P(\protect\mu )$ curves for the nonlinear tetramer modes,
originating from $\protect\mu _{5,6}^{(0)}$ in Eq. (\protect\ref{evs4}), for
different values of $\protect\lambda $, and $\protect\kappa =0.5$.}
\label{Pm4mer}
\end{figure}

\begin{figure}[tbp]
\centering
\includegraphics[width=14cm]{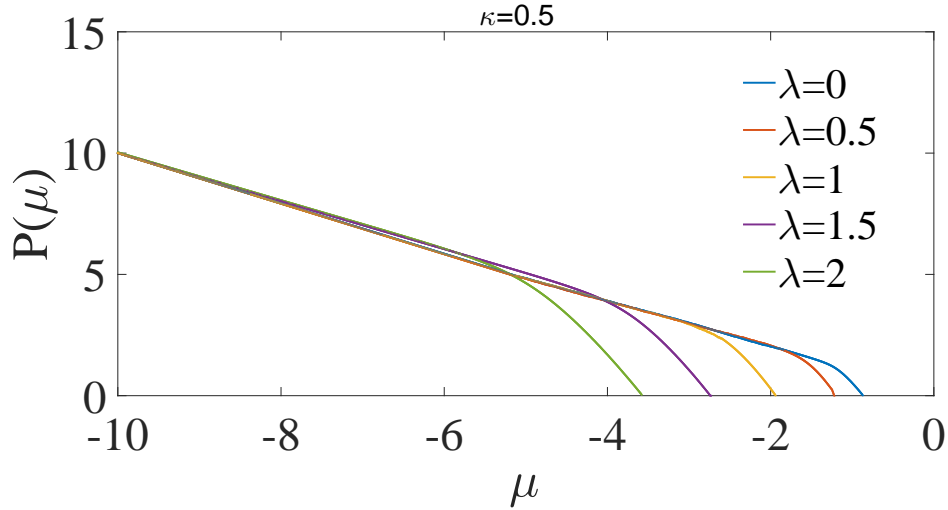}
\caption{$P(\protect\mu )$ curves for families of the nonlinear pentamer
states, originating, in the linear limit, from egenmodes corresponding to $%
\protect\mu _{9,10}^{(0)}$ in Eq. (\protect\ref{evs5}), for indicated values
of $\protect\lambda $ and $\protect\kappa =0.5$.}
\label{Pm5mer}
\end{figure}

Using numerical methods, we have found other families of nonlinear states in
all the considered $N$-mers, which originate, in the linear limit, from
trimers, tetramers, and pentamer eigenmodes associated with EVs $\mu
_{5,6}^{(0)}$, $\mu _{5,6}^{(0)}$, and $\mu _{9,10}^{(0)}$, respectively
[see Eqs. (\ref{evs3}), (\ref{evs4}), and (\ref{evs5})]. As an example of
these modes, we first present, in Fig. \ref{nl3mer}, numerically generated
nonlinear solutions in the trimer system, which feature even and odd ($+$
and $-$) spinor components, respectively:%
\begin{equation}
u_{-1}^{(+)}=u_{+1}^{(+)},~u_{-1}^{(-)}=-u_{+1}^{(-)},~u_{0}^{(-)}=0.
\label{symmetry}
\end{equation}%
In general, the families of the nonlinear trimer solutions, which exist for
different values of $\lambda $ and arbitrary values of $\kappa $, are
characterized by dependencies of the total norm on the chemical potential, $%
P(\mu )$, as shown in Fig. \ref{Pm3mer}. These nonlinear trimers are stable
in their entire existence region, according to the LSA, which is confirmed
by direct simulations. In this connection, it is relevant to mention that
all $P(\mu )$ curves in Fig. \ref{Pm3mer}, as well as in Figs. \ref{Pm4mer}
and \ref{Pm5mer} displayed below for nonlinear tetramer and pentamer modes,
satisfy the well-known Vakhitov-Kolokolov (VK)\ stability criterion, $%
dP/d\mu <0$, which is a necessary condition (but, generally speaking, is not
a sufficient one -- see the situation for the tetramers presented below) for
the stability of solitons supported by attractive nonlinearities \cite{VK}.

\begin{figure}[tbp]
\centering
\includegraphics[width=14cm]{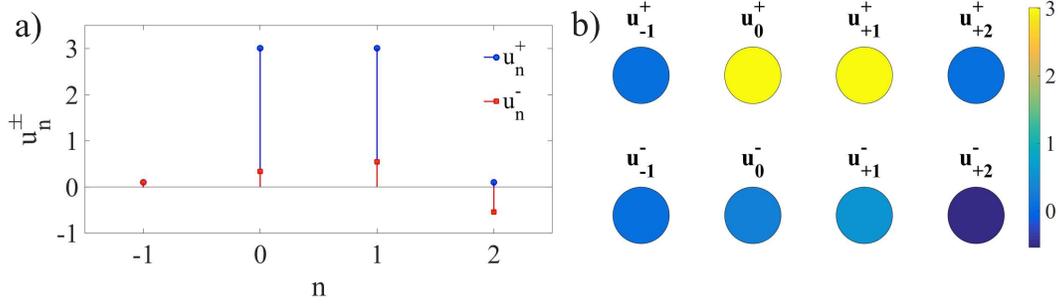}
\caption{An unstable nonlinear tetramer solution obtained at $\protect\mu %
=-10$, $\protect\kappa =0.5$, and $\protect\lambda =1$.}
\label{nl4mer}
\end{figure}

In the tetramer system, we have numerically found a family of nonlinear
solutions, which obey the symmetry relations in one spinor component,\
\begin{equation}
u_{0}^{(+)}=u_{1}^{(+)},u_{-1}^{(+)}=u_{2}^{(+)},  \label{tetra}
\end{equation}%
while the other component is asymmetric, see Fig. \ref{nl4mer}. $P(\mu )$
dependencies for the tetramer nonlinear solutions are shown in Fig. \ref%
{Pm4mer} for different values of the SO-coupling strength, $\lambda $.

\begin{figure}[tbp]
\centering
\includegraphics[width=14cm]{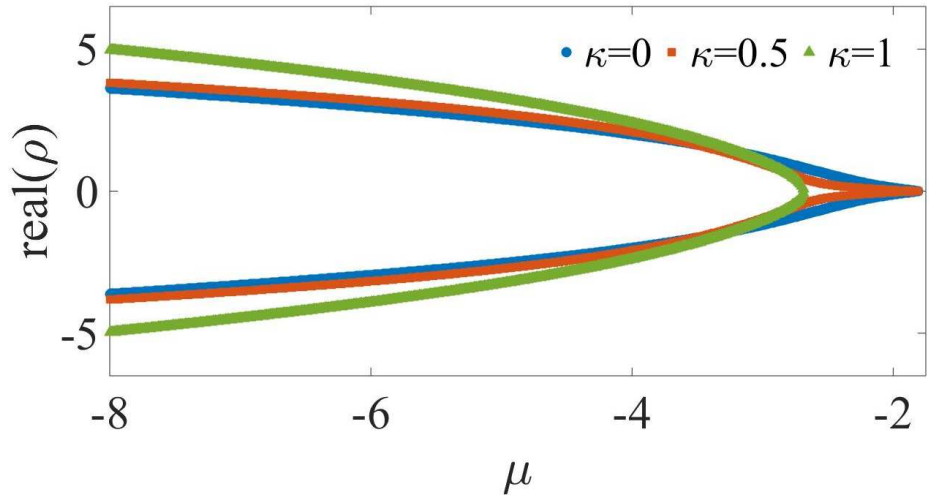}
\caption{Real parts of stability eigenvalues $\protect\rho $ for the
nonlinear tetramer modes vs. $\protect\mu $ at $\protect\kappa =0$ and $0.5$
(blue and red lines, respectively), and $\protect\kappa =1$ (the green
line), while $\protect\lambda =1$ is fixed. To avoid making the plot messy,
the zero eigenvalue for $\protect\kappa =1$ is not shown in the
corresponding stability interval, $-2.7<\protect\mu <0$. }
\label{4merEV}
\end{figure}

The LSA analysis indicates instability of the nonlinear tetramer modes in
their nearly whole existence region, in spite of the fact that the $P(\mu )$
curves satisfy the Vakhitov-Kolokolov criterion in Fig. \ref{Pm4mer}. Narrow
stability windows open only for the Manakov's type of the nonlinearity ($%
\kappa =1$). As an example, the EV spectrum of small perturbations for the
tetramer with $\lambda =1$ is displayed in Fig. \ref{4merEV} for $\kappa =0$%
, $0.5$, and $1$. A stability interval, $-2.7<\mu <0$, is found at $\kappa =1
$.

\begin{figure}[tbp]
\centering
\includegraphics[width=14cm]{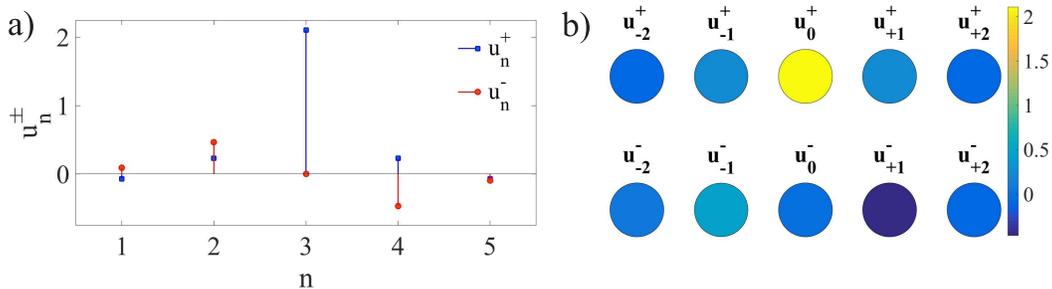}
\caption{A stable nonlinear pentamer solution numerically found for $\protect%
\mu =-5$, $\protect\kappa =0.5$, and $\protect\lambda =1$. It obeys symmetry
constraint (\protect\ref{penta}).}
\label{nl5mer}
\end{figure}

Finally, Fig. \ref{nl5mer} demonstrates that nonlinear pentamer modes,
originating, in the linear limit, from EV $\mu _{9,10}^{(0)}$ in Eq. (\ref%
{evs5}), obey the same symmetry as their trimer counterparts (cf. Eq. (\ref%
{symmetry})),%
\begin{equation}
u_{-1,-2}^{(+)}=u_{+1,+2}^{(+)},~u_{-1,-2}^{(-)}=-u_{+1,+2}^{(-)},~u_{0}^{(-)}=0.
\label{penta}
\end{equation}%
The corresponding $P(\mu )$ dependencies for certain values of $\lambda $
are displayed in Fig. \ref{Pm5mer}. According to the LSA and direct
simulations, the entire family of these nonlinear pentamer solutions is
stable.

An essential difference of the nonlinear trimer and pentamer modes
considered here from those presented in the previous sections [see Eqs. (\ref%
{Ptri}), (\ref{Ppenta}) and (\ref{Ppenta2})], is that their total norm
depends on the SO-coupling constant, $\lambda $. As a result, both types of
the nonlinear trimer and pentamer modes may coexist at the same values of
the chemical potential, $\mu $, but having different norms.

Finally, we have constructed families of linear and nonlinear $6$- and $7$-
(hexa- and septa-)mer modes. In the former case, they feature properties
similar to those of the tetramers, while in the latter case the properties
are similar to those of trimers and pentamers. Thus, the properties of
oligomers are chiefly determined by the parity of the number of their sites.

\section{Conclusion}

We have studied small linearly-shaped strings composed of two-component BEC
droplets with the intrinsic pseudo-SO (spin-orbit) coupling and attractive
nonlinearity, which can be implemented in current experiments with matter
waves trapped in optical lattices. We have found eigenmodes of these
systems, in analytical and numerical forms, and explored their stability by
means of numerical methods, through the calculation of eigenfrequencies for
small perturbations and direct simulations. The analysis has been performed
for the dimers, trimers, tetramers, and pentamers.

In the dimer system, linear-eigenmode families can be generically extended
as nonlinear modes only in the case of equal strengths of the intra- and
inter-species attractive interactions (the Manakov's nonlinearity). The
exception is the branch of the eigenmodes subject to constraint (\ref{single}%
), which exists for arbitrary values of the nonlinearity parameters. The
analysis has demonstrated that the families of nonlinear dimer modes are
stable in their whole existence regions.

In the trimer and pentamer systems, only those eigenmodes whose eigenvalues
are zero in the linear limit, $\mu =0$, can be extended, keeping their shape
unaltered, to the nonlinear system. The eigenmodes which satisfy constraint (%
\ref{single}) show properties similar to those of their dimer counterparts.
Their existence region, in the presence of the nonlinearity, is bounded by $%
\mu <0$ (negative chemical potential) and does not depend on the SO
strength, $\lambda $, while the stability boundary depends on $\lambda $. In
all cases, the stability area of the nonlinearly continued eigenmodes
shrinks with the increase of $\lambda $. On the contrary, in the tetramer
system the nonlinear modes cannot be found in the exact form, because the
respective linear spectrum does not include $\mu =0$.

In addition, in the trimer and pentamer systems the families of completely
stable nonlinear modes are found (in the numerical form), which in the
linear limit, originate from eigenmodes corresponding to $\mu \neq 0$. In
the tetramer system, such nonlinear modes are found too. However, in the
latter case those modes can be stable, in a narrow range of the system's
parameters, solely with the Manakov's type of the nonlinearity.

As concerns possibilities for extension of the analysis, one of them is to
analyze a possible supersolid phase in the truncated lattice, as suggested
by the recent experiment performed in the SO-coupled BEC\ \cite{kett}.

\section*{Acknowledgments}

The authors acknowledge support from the Ministry of Education, Science and
Technological Development of Republic of Serbia (project III45010). The work
of B.A.M. was supported, in part, by grant No. 2015616 from the joint
program in physics between the NSF and Binational (US-Israel) Science
Foundation, and by grant No. 1286/17 from the Israel Science Foundation.

\end{document}